\documentstyle[12pt]{article}

\global\arraycolsep=1pt
\begin{document}
\newcommand{\bx}{{\bf x}}
\newcommand{\tbx}{{\tilde \bx}}
\newcommand{\bu}{{\bar u}}
\newcommand{\bv}{{\bar v}}
\newcommand{\bz}{{\bf z}}
\newcommand{\bn}{{\bf n}}
\newcommand{\bm}{{\bf m}}

\newcommand{\al}{\alpha}
\newcommand{\bal}{{\bar \alpha}}
\newcommand{\be}{\beta}
\newcommand{\bbe}{{\bar \beta}}
\newcommand{\gm}{\gamma}
\newcommand{\dl} {\delta}
\newcommand{\bdl}{{\bar \dl}}
\newcommand{\ep}{\epsilon}
\newcommand{\kp} {\kappa}
\newcommand{\lm} {\lambda}

\newcommand{\bj}{{\bar J}}
\newcommand{\tj}{{\tilde J}}
\newcommand{\cc}{f_{\pi}}
\newcommand{\w}{\wedge}
\newcommand{\pr}{\partial}
\newcommand{\hf} {{1 \over 2}}
{\baselineskip=14pt 
\rightline{
 \vbox{\hbox{YITP-97-15}
       \hbox{hep-th/9704010} }}}

\vskip 10mm
\begin{center}
{\Large \bf Higher-dimensional WZW Model \\ 
              on K\"ahler Manifold and Toroidal Lie Algebra} 
                  
\vskip 12mm
{\bf
Takeo Inami, \ Hiroaki Kanno${}^{\ast}$ and \ Tatsuya Ueno}
\\ 

\vskip 15mm
\it{Yukawa Institute for Theoretical Physics}\\
\it{Kyoto University, Kyoto 606-01, Japan}\\
\it{ and }\\
\it{Department of Mathematics, Faculty of Science${}^{\ast}$}\\
\it{Hiroshima University, Higashi-Hiroshima 739, Japan} 

\end{center}

\vskip 15mm
\begin{center}
{\large \bf ABSTRACT}
\end{center}
We construct a generalization of the two-dimensional 
Wess-Zumino-Witten model on a $2n$-dimensional K\"ahler 
manifold as a group-valued non-linear sigma model with 
an anomaly term containing the K\"ahler form.
The model is shown to have an infinite-dimensional symmetry 
which generates an $n$-toroidal Lie algebra.
The classical equation of motion turns out to be 
the Donaldson-Uhlenbeck-Yau equation, which is a $2n$-dimensional
generalization of the self-dual Yang-Mills equation.
\thispagestyle{empty}
\newpage
\section{Introduction} %

There has recently been a renewed interest in field theories and 
supergravity theories in diverse dimensions.
They are believed to play the role of effective field theories of 
the conjectured ultimate theory of unification, e.g. some sort of 
superstring or supermembrane theory.
Higher-dimensional field theories also deserve studying on their 
own right. 
It is an interesting question to find the integrable structure of the 
theories, which makes them solvable and leads to some good properties,
e.g. their fine ultraviolet behaviour. 
Apparently the ultraviolet behaviour becomes worse as the space-time 
dimension increases, unless miraculous cancellation of divergences 
occurs due to large symmetry of the theories.
\par

In the past, various attempts have been pursued to construct 
higher-dimensional integrable field theories \cite{PC}, albeit without 
much success.
One approach to find such theories is to generalize 2D integrable field 
theories, respecting their good properties on the integrability.
Recently the study of a 4D analogue of  the 2D Wess-Zumino-Witten 
(WZW) model has revealed a few remarkable properties of the model 
\cite{NS},\cite{LMNS}.
The model, which we refer to as 4D K\"ahler WZW (KWZW) model,
has an infinite-dimensional (anti-)holomorphic symmetry \cite{NS} and 
is solvable in its $\it algebraic$ sector \cite{LMNS}.
It has also been shown that the model is one-loop on-shell finite in 
spite of the apparent non-renormalizability by power counting 
\cite{LMNS},\cite{K}. 
\par

In \cite{IKUX}, we have clarified that the infinite-dimensional symmetry 
generates a current algebra identified with a two-toroidal Lie algebra, 
a central extension of the two-loop algebra, which a few mathematicians 
have recently begun to study as a possible extension of the affine Kac-Moody 
algebra \cite{tta}.
It is interesting and natural to ask whether a higher-dimensional 
generalization of the 2D WZW model extends to general $2n$ dimensions.
\par

In this paper, we will construct a $2n$-dimensional analogue of the
WZW model, which we call $2n$D KWZW model since it is defined on a 
K\"ahler manifold.
The model is a group-valued non-linear sigma model (NLSM) with an addition 
of an anomaly term containing the $(n-1)$-th power of the K\"ahler form
$\omega$.
We will show that the $2n$D KWZW model has a higher-dimensional 
generalization of many properties of the 2D WZW model, e.g. it has 
an infinite-dimensional symmetry generating an $n$-toroidal Lie algebra.
The classical equation of motion of the model turns out to be the 
Donaldson-Uhlenbeck-Yau equation, which suggests a close relationship of
the model to complex and algebraic geometry. 
\par

\section{$2n$D KWZW model }  %
Let us consider a NLSM with torsion defined on a $2n$D Riemannian 
manifold $X_{2n}$. 
The basic field in the model is a mapping $\phi^a(x)$ from $X_{2n}$ to 
a target space $\cal T$ with a metric $g_{ab}$ and a two-form
$b=b_{ab}\, d\phi^a \w d\phi^b$.
The most general form of the action which contains only second-order
derivatives of $\phi^a(x)$ and is invariant under the
reparametrizations of $X_{2n}$ and $\cal T$, is written as, 
in differential form, 
\begin{equation}
S ={\cc^{2n-2} \over 4\pi} 
\int_{X_{2n}}d\phi^a \w \ast d\phi^b\, g_{ab}(\phi)
+{\gm \over 4\pi} 
\int_{X_{2n}}\Theta_{2(n-1)} \w d\phi^a \w d\phi^b\, b_{ab}(\phi) \ ,
                                                  \label{eq: act}
\end{equation}
where $\Theta_{2(n-1)}$ is a $2(n-1)$-form on $X_{2n}$.
Our convention is that the coupling constant $\cc$ has the mass
dimension $+1$, while all differential forms and the parameter $\gm$ 
are dimensionless.
\par

Let us assume the space-time $X_{2n}$ is a K\"ahler manifold and 
$\Theta_{2(n-1)}$ is the $(n-1)$-ple product of a closed K\"ahler form
$\omega$,
\begin{equation}
\Theta_{2(n-1)} = \omega^{n-1}, 
\ \ \ \ \ \ \ \ \ 
\omega = {i \over 2} \cc^2 \, h_{\al {\bar \be}} \, 
dz^{\al} \w d{\bar z}^{\be} \ ,                  \label{eq: kf}
\end{equation}
where $h_{\al \bbe}$ $(\al, \bbe = 1,2)$ is a K\"ahler metric on
$X_{2n}$. 
The target space $\cal T$ is assumed to be a group manifold 
$G$.
Then vielveins $V^i_a(\phi)$ defined from 
$g_{ab}= V^i_a V^j_b \dl_{ij}$ obey the Maurer-Cartan equation,
\begin{equation}
 \pr_a V^i_b - \pr_b V^i_a -2 f^{ijk}V^j_a V^k_b = 0 \ .
\end{equation}
This pure-gauge condition gives 
$V^i_a d\phi^a T^i = -{i \over 2}g^{-1}dg$, with generators $T^i$ of
the Lie algebra associated with the group $G$; $[T^i, T^j] = if^{ijk}T^k$, 
${\rm Tr}\,T^iT^j = \hf \dl^{ij}$.
Using it and (\ref{eq: kf}), the first term in the action 
(\ref{eq: act}) is written as
\begin{equation}
 \cc^{2n-2} \int_{X_{2n}}d\phi^a \w \ast d\phi^b\, g_{ab}
= - {2^{(n-1)} i \over (n-1)!} \int_{X_{2n}} 
\omega^{n-1}\w {\rm Tr} g^{-1}\pr g \w g^{-1}{\bar \pr}g \ ,
\end{equation}
where $\pr = (\pr/\pr z^{\al}) dz^{\al}$ 
(${\bar \pr}= (\pr/\pr {\bar z}^{\be}) d{\bar z}^{\be}$).
To re-express the second term in (\ref{eq: act}), we introduce 
a $(2n+1)$D manifold $X_{2n+1} = X_{2n} \times [0,1]$,
\begin{eqnarray}
\int_{X_{2n}}\omega^{n-1}\w d\phi^a \w d\phi^b\, b_{ab}
&&={2 \over 3} \int_{X_{2n+1}}
\omega^{n-1} \w d\phi^a \w d\phi^b \w d\phi^c H_{abc} 
\nonumber \\
&&= {1 \over 3} \int_{X_{2n+1}} 
\omega^{n-1} \w {\rm Tr}(g^{-1}dg)^3  \ ,
\end{eqnarray}
where $H_{abc}= {3 \over 2}\pr_{[a}b_{bc]}$ is the torsion on 
the manifold $G$ and is expressed by the structure constant 
$f_{ijk}$ as $H_{abc} = V^i_a V^j_b V^k_c f_{ijk}$. 
\par

We set the value of the parameter $\gm$ as 
\begin{equation}
\gm = {2^{n-1} i \over (n-1)!} \ ,               \label{eq: gm}
\end{equation}
for which a symmetry enhancement in the model takes place, as will
be shown below.
This value will also appear in the one-loop finiteness condition 
of the model.
Finally the action (\ref{eq: act}) becomes, after re-scaling 
$\cc^{2n-2} \rightarrow i \gm^{-1}\cc^{2n-2}$,
\begin{equation}
S = 
- {i \over 4\pi} \int_{X_{2n}}\omega^{n-1} \w 
{\rm Tr} g^{-1}\pr g \w g^{-1}{\bar \pr}g 
+{i \over 12\pi} \int_{X_{2n+1}} \omega^{n-1} 
\w {\rm Tr}(g^{-1}dg)^3 \ .                    \label{eq: kwzw}
\end{equation}
\par

The choice of the $\gm$ in (\ref{eq: gm}) leads to the following
identity, which is a $2n$D analogue of the Polyakov-Wiegmann (PW)
formula \cite{PW}, 
\begin{equation}
 S[gh] = S[g] + S[h] - {i \over 2\pi}
 \int_{X_{2n}}\omega^{n-1} \w {\rm Tr}g^{-1}\pr g {\bar \pr}h h^{-1} \ . 
\end{equation}
{}From this formula, we see that the action is invariant under holomorphic 
right and anti-holomorphic left symmetries, 
$g \rightarrow h_L({\bar z}^{\be}) g h_R(z^{\al})$. 
\par

The equation of motion is given by 
\begin{equation}      
 {\bar \pr} ( \omega^{n-1} \w g^{-1} \pr g) = 0 \ ,   \label{eq: em1}
\end{equation}
or equivalently,
\begin{equation}
 \pr (\omega^{n-1} \w {\bar \pr} g \, g^{-1}) = 0 \ . \label{eq: em2}
\end{equation}
{}From this equation of motion, we can identify conserved currents 
$J$ ($\bj$) corresponding to the above right (left) action symmetry as 
\begin{equation}
J =-{i \over 2\pi} \omega^{n-1}\w g^{-1} \pr g \ , 
\ \ \ \ \ \    
\bj = {i \over 2\pi} \omega^{n-1} \w {\bar \pr} g \, g^{-1} \ .
                                                  \label{eq: jbj} 
\end{equation}
\par

\section{Donaldson-Uhlenbeck-Yau equation}%

We will show that the equation of motion (\ref{eq: em1}) or 
(\ref{eq: em2}) is equivalent to the Donaldson-Uhlenbeck-Yau 
(DUY) equation in a particular gauge.
On a $2n$D K\"ahler manifold, the curvature two-form $F$ of a gauge
field is decomposed into $(2,0), (1,1)$ and $(0,2)$ components.
Then the equations 
\begin{eqnarray}
&&F_{\al\be} = F_{\bal \bbe} =0~,        \label{eq: holom} \\
&&h^{\al\bbe} F_{\al\bbe} = 0~,          \label{eq: stable} 
\end{eqnarray} 
are called the DUY equation \cite{DUY}.
Note that these equations can be derived from the action of 
the $(2n+1)$D K\"ahler Chern-Simons theory \cite{NS}.
In algebraic geometry the moduli space of stable 
holomorphic vector bundles plays a significant role.
The DUY equation provides a description of this moduli
problem in the differential geometric language.
\par

The first condition (\ref{eq: holom}) means that the connection is 
holomorphic and the second (\ref{eq: stable}) is equivalent to the 
stability of the holomorphic bundle in algebraic geometry \cite{DUY}. 
The DUY equation also appears in the Calabi-Yau compactification of 
the heterotic string theory \cite{GSW}. 
To keep $N=1$ supersymmetry after compactification to four dimensions, 
the connection of the gauge bundle over the internal Calabi-Yau space 
should satisfy the DUY equation.
\par

The holomorphy condition (\ref{eq: holom}) is solved by setting
\begin{equation}
A_{\al} = {h_L}^{-1} \pr_{\al} h_L~, 
\ \ \ \ \ 
A_{\bbe} = {h_R} \partial_{\bbe} {h_R}^{-1}~,
\end{equation}
where $h_L$ and $h_R$ are $G^{\bf C}$-valued fields.
We have complexified the gauge group and do not assume the 
hermiticity of the gauge field for a moment.
Then introducing $g \equiv h_L h_R$, we can express the $(1,1)$
component of the curvature as
\begin{equation}
F^{(1,1)} = h_R {\bar \pr} ( g^{-1} \pr g ) {h_R}^{-1}  
               = - {h_L}^{-1} \pr ({\bar \pr} g \, g^{-1} ) h_L~.
\end{equation}
Since the stability condition (\ref{eq: stable}) can be written as 
\begin{equation}
\omega^{n-1} \wedge F^{(1,1)} ~=~ 0~,
\end{equation}
we see that it reduces to the equation of motion of the KWZW model
(\ref{eq: em1}) or (\ref{eq: em2}).
Thus the $2n$D KWZW model gives the $2n$D DUY equation and hence 
classically describes the moduli space of stable holomorphic vector 
bundles over a $2n$D K\"ahler manifold. 
\par

In four dimensions, the DUY equation reduces to the anti-self-duality
(ASD) condition on a K\"ahler manifold. 
Hence the equation of motion of the 4D KWZW model is nothing but 
the ASD condition, which has been discussed by several authors.
As we have seen above, the DUY equation is a natural 
higher-dimensional analogue of the ASD condition specific to
four dimensions. 
More than a decade ago, Corrigan et al. proposed the following 
first-order system for a gauge field \cite{CDFN},
\begin{equation}
F_{\mu\nu} ~=~\hf T_{\mu\nu\rho\sigma} F^{\rho\sigma}~, 
                                                     \label{eq: hinst}
\end{equation}
as a higher-dimensional (anti-)self-duality condition, or the instanton 
equation. 
\par
 
In the dimension $D=4$, we can take $SO(4)$ invariant tensor 
$\ep_{\mu\nu\rho\sigma}$ as a totally anti-symmetric fourth rank 
tensor $T_{\mu\nu\rho\sigma}$, which gives the well-known instanton 
equation.
If $D > 4$, $T_{\mu\nu\rho\sigma}$ is no longer invariant under $SO(D)$, 
a holonomy group of a generic orientable Riemannian manifold.
However, if the holonomy group is reduced to a subgroup $G \subset SO(D)$,
we have a chance to obtain a $G$-invariant tensor $T_{\mu\nu\rho\sigma}$ 
and to define an analogue of the instanton equation (\ref{eq: hinst}).
This is the case for a $2n$D K\"ahler manifold with the holonomy $U(n)$.
For example, on a 6D K\"ahler manifold with complex coordinates 
($z^{\al}, {\bar z}^{\al}$), the equation (\ref{eq: hinst}) with 
the tensor
\begin{equation}
T_{\mu\nu\rho\sigma}~=~\ep_{\mu\nu\rho\sigma 1 {\bar 1} }
 + \ep_{\mu\nu\rho\sigma 2 {\bar 2} }
 + \ep_{\mu\nu\rho\sigma 3 {\bar 3} } \ ,
\end{equation}
gives the DUY equation, that is, the equation of motion of the 6D KWZW 
model.
\par

The classical integrability of the ASD equation can be seen by
representing it in the Lax form.  
Possible higher-dimensional generalization of this first-order 
formulation of the ASD equation was considered by Ward \cite{W}.
Generally the DUY equation does not coincide with equations in \cite{W} 
and does not seem to allow a Lax representation. 
However, in the next section, we will show that the KWZW model has an 
infinite-dimensional current algebra. 
We believe that the algebra suggests a kind of solvability in some 
sector of the model.

\section{n-toroidal Lie algebra} %

To obtain a current algebra in the model, we start with the 
classical Poisson bracket (P.B.) in the light-cone frame. 
The P.B. is defined by finding the symplectic structure of 
the model.
In the following, we assume that the space-time $X_{2n}$ is flat;
$h_{\al \bal} = \hf$, ($\al = 1,\cdots,n$) with the others zero,
and use the notation $u = z^1,\ v^a = z^{a+1}$ 
$(\bu = {\bar z}^1,\ \bv^a = {\bar z}^{a+1})$, ($a = 1,\cdots,n-1$).
We take $\bu$ as our time coordinate, while the 
space variables are denoted as $\bx = (u, v^a, \bv^a)$.
In this light-cone frame, the action (\ref{eq: kwzw}) is first-order 
in time derivatives, that is, it is already in Hamiltonian form. 
Therefore, we can easily obtain the symplectic two-form 
$\Omega_{ij}(\bx,\bx')$ from the action,
\begin{equation}
\Omega_{ij}(\bx,\bx') = {\kp \over 2^n} \dl_{ij}
                        \pr_u \dl(u-u') \dl^{2(n-1)}(v-v') \ ,
\end{equation}
where $\kp = {i^{n^2} \over 2^n\pi}\cc^{2(n-1)}(n-1)!$ and 
$\dl^{2(n-1)} (v - v') = \displaystyle{\prod_{a=1}^{n-1}} 
\dl (v^a - v^{\prime a}) \dl (\bv^a -\bv^{\prime a})$.
Henceforth, we take the values of the coordinates real by using 
an analytic continuation.
The P.B. is defined using $\Omega^{ij}$, the inverse of the symplectic 
form,
\begin{eqnarray}
[ f_1(\bx), f_2(\bx') ]_{\rm P.B.}  
&&= \int d \tbx d \tbx'  
\, \dl f_1(\bx) /(\dl g g^{-1}(\tbx))^i \, \Omega^{ij}(\tbx,\tbx') \,
\dl f_2(\bx') /(\dl g g^{-1}(\tbx'))^j 
\nonumber \\
&&= 2^{n+1} \kp^{-1} \int d\tbx 
{\rm Tr}(\dl f_1(\bx)/\dl g g^{-1}(\tbx) \pr^{-1}_{\tilde u} 
\dl f_2(\bx') /\dl g g^{-1}(\tbx))    \ .         
\end{eqnarray}
\par

The components of the current $J$ are given from (\ref{eq: jbj}) as 
$J_A = \kp g^{-1} \pr_A g$, ($A = u, v^a$).
Since $\bu$ is our time coordinate, from the conservation laws, 
$J^i_u$ are generators of the holomorphic right action symmetry 
$g \rightarrow g h_R(z^{\al})$.
If we take the $u$ as the time coordinate instead, then 
$\bj^i_{\bu}$ become generators of another anti-holomorphic 
left action symmetry. 
The P.B.'s of currents are
\begin{eqnarray}
[{\rm Tr}&&(XJ_A)(\bx), {\rm Tr}(YJ_B)(\bx')]_{\rm P.B.}
\nonumber \\
&&= - 2^{n-1}\kp \, 
  {\rm Tr}\{gXg^{-1}(\bx) \pr_A \pr^{-1}_u \pr_B (gYg^{-1}(\bx)
  \dl(u-u') \dl^{2(n-1)}(v-v'))\} \ ,                   
\end{eqnarray}
where $X$ and $Y$ are elements of the algebra associated with 
the group $G$.
In particular, for $(A,B)=(u,u)$,
\begin{equation}
[J^i_u(\bx), J^j_u(\bx')]_{\rm P.B.}
= 2^n i f_{ijk}J^k_u(\bx) \dl(\bx-\bx') - 2^n \kp \dl^{ij}\pr_u
\dl(\bx-\bx')  \ .                            \label{eq: ca}
\end{equation}
\par

The current algebra (\ref{eq: ca}) can be cast into a more familiar form 
by making the mode expansion.
To this end, we compactify the space directions such that $u$, $v^a$ and 
$\bv^a$ take values in $[0, 2\pi]$.
The charge of the holomorphic right action transformation by the 
generators $J$ is then given by  
\begin{equation}
Q^k = \int^{2\pi}_0 du \prod_{a=1}^{n-1}dv^a \ep_k(u,v) \tj^k \ ,
\ \ \ \ \
\tj^k ={1 \over 2^n}\int^{2\pi}_0 \prod_{a=1}^{n-1}d\bv^a J^k_u(\bx) 
\ .                                           \label{eq: qtj}
\end{equation}
{}From the conservation laws, we see that $\tj^k$ are functions of 
$u$ and $v^a$ only; $\tj^k = \tj^k(u,v^a)$. 
We use the notation $\bz = (u,v^a)$. 
Then the P.B.'s for $\tj^i(\bz)$ are written as
\begin{equation}
[\tj^i(\bz), \tj^j(\bz')]_{\rm P.B.} 
= i f_{ijk} \tj^k(\bz) \dl(\bz-\bz') 
- {(2\pi)^{n-1} \over 2^n} \kp \dl^{ij} \pr_u \dl(\bz-\bz') \ .  
                                              \label{eq: tj}
\end{equation}
\par

The parameter functions $\ep_k(\bz)$ are defined on the $n$-torus 
and can be expanded as 
\begin{equation}
\ep_k(\bz) = \sum_{\bm} \ep^{\bm}_k \, e^{i\bm \cdot \bz} \ ,
\end{equation}
where $\bm = (m_1,m_2,\cdots,m_n) \in {\bf Z}^n$ and 
$\bm \cdot \bz = m_1u + m_2v^1+ \cdots + m_n v^{n-1}$. 
Correspondingly, the modes of the currents $\tj^k(\bz)$ are defined as 
\begin{equation}
Q^k = \sum_{\bm} \ep^{\bm}_k J^k_{\bm} \ ,
\ \ \ \ \ 
J^k_{\bm} = \int_0^{2\pi} du \prod_{a=1}^{n-1}dv^a 
e^{i\bm \cdot \bz} \tj^k(\bz)\ .
\end{equation}
Then from (\ref{eq: tj}) we obtain
\begin{equation}
[J^i_{\bn}, J^j_{\bm}]_{\rm P.B.} = i f_{ijk} J^k_{\bn+\bm} 
 + i \lm_1 n_1 \dl^{ij}\dl_{\bn, -\bm} \ , 
\ \ \ \ \ \ \ \   \lm_1 = {(2\pi)^{2n-1} \over 2^n} \kp \ . 
                                                 \label{eq: nt}
\end{equation}
This P.B. algebra is an infinite-dimensional 
(i.e.\,$\bm \in {\bf Z}^n$) Lie algebra with central extension, 
which is called $n$-toroidal Lie algebra \cite{tta}.
Note that the integration over $\bv^a$ of the original currents
$J^i_u(\bx)$ is performed in (\ref{eq: qtj}) to obtain the 
generators $J^i_{\bn}$ of the toroidal algebra.
This $(n-1)$-ple integration also ensures the decouping of holomorphic 
and anti-holomorphic sectors of the total algebra; 
$[J^i_{\bn}, \bj^j_{\bm}]_{\rm P.B.}= 0$.
\par

There appears only one center proportional to $\lm_1 n_1$ in 
the P.B. (\ref{eq: nt}), which corresponds to the coordinate $u$.
The reason for this is that we have chosen $\bu$ as the time
coordinate and defined the P.B. with respect to it. 
It is, however, known that the $n(\geq 2)$-toroidal Lie algebra is an 
infinite-dimensional central extension of the $n$-loop algebra
unlike the $n=1$ (affine Kac-Moody) case \cite{tta}.
Hence, other central terms may arise in (\ref{eq: nt}) when we 
evaluate quantum corrections to the P.B. algebra. 
\par
 
\section{Discussion}  %

We have shown that the $2n$D KWZW model gives the 
Donaldson-Uhlenbeck-Yau (DUY) equation, or equivalently 
a generalization of the anti-self-dual (ASD) condition, as 
the equation of motion and that the model has an 
infinite-dimensional symmetry generating the $n$-toroidal 
Lie algebra. 
A natural question is whether the symmetry leads to nice 
properties of the model, such as the integrability and the 
one-loop finiteness.

The equation of motion in the 4D case reduces to the ASD 
Yang-Mills equation.
It has been argued that behind the ASD equation there exists 
another type of infinite-dimensional symmetry of the loop algebra
\cite{DC}, which is different from our toroidal algebra symmety 
in a few points.  
One crucial difference is that the former gives an affine Kac-Moody 
algebra with no central extension, whereas our toroidal algebra has a 
central term, which is an inevitable consequence of the anomaly 
term in the KWZW model.
In this sense, the former is an analogue of the (hidden) symmetry of
the 2D principal chiral model, while our symmetry is a genuine
extension of the affine Kac-Moody symmetry in the 2D WZW model.
Another difference is that the toroidal algebra symmetry is valid not 
only at the level of equation of motion but also at the level of 
action.  
This is a great advantage in an attempt to quantizing the model.
\par

The above good properties of the toroidal algebra symmetry hold in 
general $2n$-dimensions.
It is worth investigating the integrability of the KWZW model and 
clarifying the role of the $n$-toroidal algebra.
\par

In four dimensions, the one-loop ultra-violet divergences, which 
haunt the NLSM, have been shown to cancel completely, partially due 
to the two-toroidal algebra symmetry.
Whether the same cancellation of the divergences persists in general 
$2n$D KWZW model is an interesting issue.
We are currently investigating the one-loop ultra-violet behaviour of 
the model in six and eight dimensions. 
We have witnessed cancellation of many divergent terms, but have not 
completed to check whether all such terms are eliminated or not.
\par

The 4D KWZW model appears in an approach to the theory of unification 
based on the N=(2,1) heterotic string \cite{KMO}.
There, the (anti-)self-dual gauge field is a mapping (M-brane) from
a (2+2)D world-volume to a (10+2)D target space, and various
superstrings and supermembrane are obtained by reducing the (2+2)D to 
(1+1)D and (2+1)D, respectively. 
The KWZW model describes the world-volume dynamics of the M-brane
and may play the same role as conformal field theory in string theory. 
\par

\vskip 20mm

We are grateful to Yoshihisa Saito for useful discussions on toroidal 
Lie algebras. 
We also wish to thank Chuan-Sheng Xiong, who collaborated with us in 
the early stage of the present work.
This work is supported partially by Grant in Aid of the Ministry of 
Education, Science and Culture.
T. U. is supported by the Japan Society for the Promotion of Science, 
No. 6293.
\newpage

\end{document}